# Terahertz cyclotron emission from two-dimensional Dirac fermions

S. Gebert[1,2][‡], C. Consejo[1][‡], S.S. Krishtopenko[1], S. Ruffenach[1], M. Szola[1], J. Torres[2], C. Bray[1], B. Jouault[1], M. Orlita[3,4], X. Baudry[5], P. Ballet[5], S.V. Morozov[6,7], V.I. Gavrilenko[6,7], N.N. Mikhailov[8,9], S.A. Dvoretskii[8,10], F. Teppe[1][*]

[1] Laboratoire Charles Coulomb (L2C), UMR 5221 CNRS-Université de Montpellier, F-34095 Montpellier, France

[2] Institut d'Electronique et des Systèmes (IES), UMR 5214 CNRS, Université de Montpellier, Montpellier 34095, France

[3] Laboratoire National des Champs Magnétiques Intenses, CNRS-UGA-UPS-INSA-EMFL, Grenoble, France

[4] Institute of Physics, Charles University, CZ-12116 Prague, Czech Republic

[5] CEA, LETI, MINATEC Campus, DOPT, 17 rue des martyrs 38054 Grenoble Cedex 9, France

[6] Institute for Physics of Microstructures of Russian Academy of Sciences, 603950, Nizhny Novgorod, Russia

[7] Lobachevsky State University of Nizhny Novgorod, 603950, Nizhny Novgorod, Russia

[8] A.V. Rzhanov Institute of Semiconductor Physics, Siberian Branch of Russian Academy of Sciences, 630090 Novosibirsk, Russia

[9] Novosibirsk State University, 630090, Novosibirsk, Russia

[10] Tomsk State University, Tomsk, Russia

[‡] These authors contributed equally
[*] Corresponding author

Since the emergence of graphene, we have seen several proposals for the realization of Landau lasers tunable over the terahertz frequency range. The hope was that the non-equidistance of the Landau levels from Dirac fermions would suppress the harmful non-radiative Auger recombination. Unfortunately, even with this non-equidistance an unfavorable non-radiative process persists in Landau-quantized graphene, and so far no cyclotron emission from Dirac fermions has been reported. One way to eliminate this last non-radiative process is to sufficiently modify the dispersion of the Landau levels by opening a small gap in the linear band structure. A proven example of such gapped graphene-like materials are HgTe quantum wells close to the topological phase transition. In this work, we experimentally demonstrate Landau emission from Dirac fermions in such HgTe quantum wells, where the emission is tunable by both the magnetic field and the carrier concentration. Consequently, these results represent an advance in the realization of terahertz Landau lasers tunable by magnetic field and gate-voltage.

## Introduction

The cyclotron or Landau level emission occurs when non-equilibrium electrons, which are subject to an external magnetic field, recombine radiatively between two magnetic quantum levels, so-called Landau levels (LL). Considering a parabolic energy-momentum dispersion and the accumulation of all electrons in a single LL, $n$, the emission rate ($n \rightarrow n-1$) becomes proportional to $n$, while the absorption rate ($n \rightarrow n+1$) is proportional to $n+1$. Thus, in systems with parabolic dispersion, absorption always outweighs amplification. When, on the other hand, these electrons have relativistic energies, i.e. in systems with non-parabolic energy-momentum dispersion, the LLs become non-equidistant in energy, which makes it possible to overcome absorption losses. The dispersion non-parabolicity is therefore a crucial factor for the development of cyclotron resonance (CR) sources. As a matter of fact, this emission process has for decades been the subject of a comprehensive study in the fields of plasma physics [1], vacuum electronics, and relativistic electronics for the development of electron cyclotron masers [2,3] and free-electron lasers [4]. In parallel, and despite the fact that transitions between LLs are strongly affected by non-radiative processes [5], particular attention has been paid to semiconductors, in which the effective mass of charge carriers is small compared to the one of the free electrons, and thus, the splitting of LLs is much greater than in vacuum.

More recently, graphene and its linear band structure was also proposed as an interesting candidate towards a lossless Landau laser [6,7,8,9,10] operating with relatively low magnetic field on the principle of gyrotrons. It was supposed to have the double benefit of reducing non-radiative Auger scattering, given the non-equidistance of its LLs, and of presenting an extreme tunability of the cyclotron radiation frequency in weak magnetic field, due to the low cyclotron mass of Dirac fermions (DFs)[11]. This idea has been successfully verified recently in a gapless HgCdTe bulk crystal and interpreted as due to the strict non-equidistance of the LLs [12]. However, the population inversion of two-dimensional (2D) Dirac electrons in Landau quantized graphene had still not been observed until now. On the contrary, very fast dynamics was clearly observed during the resonant excitation of optical transitions between low-energy LLs in graphene under strong magnetic field [13]. Despite this LL non-equidistance, Auger scattering was identified as the predominant mechanism for carrier redistribution in Landau quantized graphene. Indeed, due to the square root dependence of the energy on the Landau level index, one can still find transitions that fulfill the energy conservation law, opening up the channel for Auger scattering [14,15]. This seemed to negate the chances of observing a cyclotron emission in graphene and any other 2D system with a conical band structure described by the Dirac Hamiltonian.

However, the situation is different for DFs in semiconductor quantum wells (QWs) (like HgTe/CdHgTe QWs [16,17] and InAs/Ga(In)Sb QWs [18,19]), whose specificities make it possible to overcome non-radiative Auger recombination. Here, the Dirac Hamiltonian describing the low-energy band structure of these QWs [20] has additional terms absent in graphene. These terms not only transform a trivial insulator into a topological insulator (having an inverted band-gap), but also break equidistant LL subsets, which suppress Auger recombination for massless DFs. The latter makes semiconductor QWs particularly suitable candidates for gyrotron-like LL lasers with the highest tunability in the THz frequency range.

In this work, we report on the observation of electrically-driven cyclotron emission of 2D massive and massless DFs in HgTe/CdHgTe QWs due to the suppression of Auger scattering. Our experimental results show a continuous tunability of the emission frequency in magnetic field and in carrier concentration, over the range from 0.5 to 3 THz, thus covering the terahertz gap [21] over which a tunable laser source is still in great demand.

**Results**
*Landau levels of Dirac fermions in HgTe QWs*
Figure 1 compares the LLs of Dirac fermions in graphene and HgTe QWs. As can be seen, in graphene (Fig. 1a), the perfect square-root dependence of the energy on the LLs indices counterintuitively opens a new channel for Auger scattering as its LLs spectrum ($E_n \propto \sqrt{n}$ for $n = 0, 1, 2,...$) includes subsets of equidistant levels [16,17]. In other words, the optical transition between LLs from $n = 0$ to $n = 1$ for instance, has the same energy as from $n = 1$ to $n = 4$, from $n = 4$ to $n = 9$, etc. Note that no equidistant LL subsets can be found for massive DFs, like relativistic electrons used in gyrotrons. The most essential difference between massless DFs in semiconductor QWs and those in graphene is the absence of the series of equidistantly spaced LLs, as existing in graphene (Fig. 1b). Therefore, undesirable non-radiative Auger recombination may be suppressed for the Landau-quantized DFs in semiconductor QWs. Moreover, the band-gap for the DFs in semiconductor QWs can be tuned-at-will by adjusting the quantum confinement [17,19]. This band-gap opening further deviates the LL energies from the typical square root behavior known in graphene (see Fig.1c), which should decrease the non-radiative Auger recombination as well. Let us now show how these combined effects make possible to observe an efficient THz Landau emission in HgTe QWs.

*THz cyclotron emission analysed by Landau spectroscopy*

To observe the cyclotron emission of Landau-quantized Dirac electrons in HgTe QWs, we have employed a Landau spectroscopy technique. The electrons in HgTe QWs are heated by electric field pulses, flattening the Fermi distribution and, thus, populating upper empty Landau levels. These higher energy electrons then recombine by spontaneously emitting a photon at the energy difference between the LLs involved. Our Landau spectrometer can be used in two different modes. The first mode (Mode 1) consists in keeping the magnetic field applied to the detector constant, to fix the detection energy, while the magnetic field applied to the sample is scanned. The second mode (Mode 2) consists on the contrary in keeping the magnetic field on the sample constant by sweeping the one applied to the detector, to directly measure the emission spectrum. In order to favor the observation of cyclotron radiation, we studied a series of HgTe QWs with a band gap in the range from 0 to 30 meV (see Supplemental Materials). All the measurements were carried out at a temperature of 4 K.

Figure 2 shows the emission spectra of two 8 nm thick HgTe QWs of different electron concentrations. All spectra contain an emission line, symmetrical in positive and negative magnetic fields, and already visible at low energies (see Fig. 2 a) and c)). The line position clearly evolves linearly with the magnetic field, allowing determining the cyclotron mass $m_c$. Note that the slope of the resonant energies as functions of magnetic field depends on the carrier concentration due to non-parabolic band structure of HgTe QWs (see Fig. 1(b,c)). Figures 2(b,d) provide the corresponding emission spectra at fixed magnetic fields for both samples. The emission peaks have a Gaussian-like shape, whose amplitudes increase first with the magnetic field and decrease after a certain energy value, which varies from one sample to another. Therefore, this non-monotonic evolution of the emission intensity cannot be attributed only to the reststrahlen band in the QWs (15–20 meV for HgTe/Cd$_x$Hg$_{1-x}$Te), nor to the detector's limits. Indeed, $n$-InSb can be used as a narrow-band and tunable detector up to about 22 meV, where its efficiency is limited by the InSb reststrahlen band [22]. Moreover, depending on the carrier concentration, the edge of the reststrahlen band can be shifted towards energies slightly lower than 15 meV and it is therefore probable that the disappearance of the emission signal beyond 12 meV is due to the vicinity of the HgTe/Cd$_x$Hg$_{1-x}$Te reststrahlen band. However, the observed decrease in emission amplitude starts at much lower energies, on the order of 4 to 6 meV. This phenomenon therefore does not seem to be linked to the influence of the reststrahlen band. Another way to understand this behaviour would be to consider that when the sample leaves the incipient Landau quantization regime for a fully resonant mode involving discrete and well-separated LLs, the electron relaxation time decreases. The similar effect was observed in graphene during the resonant excitation of optical transitions between low-energy LLs [16]. This interpretation is however unlikely since, as discussed before, it is difficult to find subsets of LLs with the same energy in HgTe QWs and as it can be seen in Figs. 3(a) and (b), the samples are far from the quantum limit. This drop in amplitude cannot be attributed neither to the non-linear magnetoresistance in these samples (see Fig. SM7 in Supplemental Materials). Therefore, we attribute this decrease to the combine actions of different effects such as the vicinity of reststrahlen band, the detector's response and the uncontrolled out-of-equilibrium carrier distribution in crossed electric and magnetic fields.

*Cyclotron emission tunable in magnetic field and concentration*
As discussed above, given the carrier mobility and the range of the applied magnetic field, the samples are in the conditions of incipient Landau quantization. The low-index LLs are well separated, but given their non-equidistance, the distribution functions of the higher index states overlap to form a quasi-continuum. In this regime, the cyclotron frequency evolves linearly with the magnetic field and the system [23] can be treated in a semi-classical manner. Applying a semi-classical quantization rule to the low-energy band dispersion, the cyclotron mass $m_c$ in the conduction band as a function of Fermi momentum $k_F = \overline{2\pi n_S}$ (where $n_S$ is an electron concentration) has the form:

$$m_c = \frac{\hbar^2 \overline{\left(M - \mathbb{B}k_F^2\right)^2 + A^2 k_F^2}}{s\left(A^2 - 2\mathbb{B}\left(M - \mathbb{B}k_F^2\right) - 2D\overline{\left(M - \mathbb{B}k_F^2\right)^2 + A^2 k_F^2}\right)}. \quad (1)$$

Here, $M$ describes the band-gap, while $A$, $\mathbb{B}$ and $\mathbb{D}$ are parameters determined by the QW width and the barrier materials (see Supplemental Materials).

Figure 3 summarizes the position of the emission line at different magnetic fields and the corresponding $m_c$ as functions of $n_S$ extracted from magneto-transport data (see Supplemental Materials). As clear from Fig. 3(c), even though $m_c$ was extracted from emission experiments with the out-of-equilibrium carrier distribution, its values are in good agreement with the ones given by Eq. (1). In order to emphasize the tunability of cyclotron radiation, Fig. 3(d) also shows the resonant energy for the two groups of QW widths at the two magnetic field values (0.5 T and 1 T). Here, we note a good agreement between the experimental values and the theoretical calculations in the energy range between 2 and 9 meV (0.5 to 2.2 THz). Additionally, $m_c$ extracted from emission results also agrees well with the results of magneto-absorption experiments performed on the same sample. Interestingly, the CR energy probed by magneto-absorption (cf. Supplementary Materials) is almost insensitive to temperature up to 100 K. The latter gives the hope to observe the Landau emission in HgTe QWs at significantly higher temperatures than in this work.

**Discussion**

Before discussing the conditions required for stimulated emission and gain, it is worth mentioning those for streaming effect and population inversion of the LLs. In the semi-classical regime, the homogeneous broadening of the CR, induced by elastic scattering on impurities and phonons, is directly related to the DC-conductivity at zero magnetic field[24]. The CR linewidth can therefore be compared with the reverse of the free carrier momentum relaxation time or total lifetime $\tau_t$. The full-width at half-maximum of the emission lines in our experiments is on the order of 3-4 meV (corresponding to a frequency $f \approx 800$ GHz and a total life time $\tau_t \approx 0.2$ ps), which is about 10 times larger than the collision broadening values $\Delta\varepsilon = \hbar/\tau = \hbar e/\mu m_c$ extracted from mobility $\mu$ measurements (see Supplementary Materials). It should also be noted that the linewidth of the emission line is twice that of the absorption. Indeed, spontaneous cyclotron emission involves hot electrons, which are distributed in a much broader energy range than the carriers probed by quasi-equilibrium magneto-absorption involving the electrons in the vicinity of Fermi energy. Therefore, the electric field $E$ yielding non-equilibrium electron distribution also influences the form of emission line in the systems with non-parabolic band structure as $m_c$ increases with the energy[25]. Additionally, the form of the emission line can be also affected by the inhomogeneous broadening due to electron-phonon scattering[26], as well the Stark broadening induced by ionized impurities[27]. The latter can be a dominant contribution into the emission linewidth as previously observed by Gornik et al.[28].

To evaluate the conditions required for streaming effect and possible population inversion of LLs, we compare the momentum relaxation time determined from the carrier mobility (see table 1 in Methods), with the time-of-flight $T_{op}$ it takes for an electron to reach the longitudinal optical (LO) phonon energy[29] (18.3 meV for HgTe QWs [[30]]). With $T_{op} = \frac{V_{op} m_c}{eE}$ and $V_{op} = \sqrt{\frac{2\hbar\omega_{op}}{m_c}}$, we get $T_{op} \approx 6$ ps with $E \approx 100$ V/cm typically used in our experiments. The momentum relaxation time determined by our transport measurements is on the order of 1–2 ps. In other words, the electrons are scattered before they reached the LO phonon energy $\hbar\omega_{op}$. In order to reach $E_{op}$ before scattering, $\mu$ should be of the order of $2\cdot10^5$ cm$^2$/V·s with an electric field from 300 to 600 V/cm. A population inversion would be possible only if the following condition were satisfied[31] $1 \leq \frac{V_{op}}{\frac{E}{B}} \leq 2$, which can be achieved, for instance, for $E \approx 2000$ V/cm and magnetic field in the range from 0.2 to 0.4 T. These values are achievable in sub-millimeter-sized grating gate devices fabricated from high-mobility HgTe QWs[32].

Let us now consider a lasing threshold in possible Landau laser based on HgTe QWs. To estimate the population inversion conditions, we assume inducing a gain comparable to that of THz p-Germanium lasers. The latter has typically a gain coefficient[33] $\gamma = \frac{N_{2D}}{d_{QW}} \lambda^2 4\pi^2 \tau_{sp} \Delta\nu$ of the order of 0.05 cm$^{-1}$ [[34,35]], where $N_{2D}$ corresponds to the population difference, $d_{QW}$ is the QW width, $\lambda$ is the wavelength of light in the medium and $\Delta\nu = 1/2\pi\tau_t$. For relativistic electrons, the spontaneous cyclotron radiative lifetime is appraised as[36]: $\tau_{sp}^{-1} = \alpha \left(\frac{v_F}{c}\right)^2 \omega_c$, where $\alpha$ is the fine structure constant, $v_F$ and $c$ are the Fermi and light velocities, respectively. In the THz frequency range, $\tau_{sp}$ is therefore on the order of 1 $\mu$s, while $\tau_t \approx 0.2$ ps as was determined above. Therefore, if one considers for instance $\lambda = 300$ μm, it is necessary to carry out a population inversion of the order of $N_{2D} \approx 1.7 \cdot 10^3$ cm$^{-2}$ for $d_{QW} = 10$ nm, which seems achievable in the pulsed excitation regime.

Finally, we point out that a Landau laser based on HgTe QWs can have much greater tunability in the quantum limit than we observed under incipient Landau quantization. Indeed, among the whole set of LLs in HgTe QW, there are two specific levels with zero indices (see Fig. 1), whose optical transitions have the highest resonant energies. Particularly, the highest resonant energy in Fig. 3(a) corresponds to one of these previously observed transitions in magnetoabsorption [17]. For this transition to be dominant in Landau emission with magnetic fields below 1 T, it is necessary to reduce $n_S$ to $10^{10}$ cm$^{-2}$, which can be done by means of a semi-transparent gate deposited on top of the HgTe QWs.

**Conclusion**
We have experimentally observed cyclotron emission of massive and massless DFs in HgTe QWs. The emission is favored both by the band-gap opening and the specific band dispersion in HgTe QWs. The latter breaks the possibility to find series of equidistantly spaced LLs and thus suppresses non-radiative Auger recombination inherent in graphene. We have demonstrated a continuous tunability of the emission frequency in the range from 0.5 to 3 THz under low magnetic fields. These results lead us to consider HgTe QWs as promising materials for the development of a THz cyclotron laser largely tunable by a weak magnetic field or even by the gate voltage using a permanent magnet.

**Additional information**
The authors declare no competing financial interests. Correspondence and requests for materials should be addressed to F.T.

**Acknowledgements**
This work was supported by the Terahertz Occitanie Platform (FT, JT), by the CNRS through IRP "TeraMIR" (FT, MO, VIG), by the French Agence Nationale pour la Recherche for Colector (ANR-19-CE30-0032) (FT, MO, VIG), Stem2D (ANR-19-CE24-0015) (JT), and Equipex+ Hybat (ANR-21 -ESRE-0026) projects (FT), by the European Union through the Flag-Era JTC 2019 - DeMeGras project (ANR-19-GRF1-0006) (FT) and the Marie-Curie grant agreement No 765426 (SG), from Horizon 2020 research and innovation program, and by the Center of Excellence «Center of Photonics» funded by The Ministry of Science and Higher Education of the Russian Federation, contract № 075-15-2022-316 (SVM, VIG). The authors would like to acknowledge Christian Lhenoret for technical support, Pr. Luca Varani for financial support, and Pr. Wojciech Knap for fruitful discussions and valuable support. FT and CC would also like to thank Dr. Stephane Bonifacie for all the passionate discussions, for his friendship and his timeless presence.

**Author contributions statement**
The experiment was proposed by FT. The samples were fabricated by NNM, SAD, XB, and PB. THz cyclotron emission experiments were carried out by SG and CC. Characterization measurements were conducted and analysed by SR, MS, CB, and BJ. SG, CC and MS handled the data and prepared the figures. FT, and SSK wrote the manuscript and SG, MO, SVM, and VIG corrected it. All co-authors discussed the experimental data and interpretation of the results.

**Competing interests statement**
The authors declare no competing financial interests.

## Methods

A part of the QWs were grown on GaAs (013) substrates with ZnTe and CdTe buffers (see Fig. SM9 a) using molecular beam epitaxy (MBE) with *in situ* ellipsometric control[36] of the layer composition and thickness[36]. This growth method has been shown to provide high-quality structures, investigated in a large number of works[36]. The other part was grown on CdTe (001) substrate with again a CdTe buffer layer. The tensile strained HgTe well and HgCdTe barrier layers are grown at the same temperature (160°C) with a constant growth rate of 1 monolayer/s. Details on the growth and material characterization can be found in [36]. HgTe/CdTe structures grown using the same procedure are associated to high mobility electronic transport with clear Dirac characteristics [36] and have also been used to demonstrate record room-temperature spin to charge conversion[36] (Fig. SM9 a). The samples were cut into squares of approximately 5x5 mm$^2$ and gold wires were soldered to indium balls placed on the sample' surface, acting as ohmic contacts, spaced approximately 1 mm apart. It is assumed that the distribution of the electric field is fairly uniform and that the energy is dissipated throughout the "volume" of our samples. Therefore, the cyclotron emission is naturally expected to come from the entire sample. The dependence of the emission amplitude on the applied voltage was measured in sample S6 (see Fig. SM6). It is clearly seen that the amplitude of the cyclotron emission increases with the applied voltage. The concentration and the mobility of the carriers were determined by magneto-transport measurements (detailed in the Supplementary Materials) whose results are grouped in Table SM2. The experimental setup is a Landau spectrometer which consists of two independent 8 T and 14 T superconducting coils placed in the same cryostat (Fig. SM9 b). The sample to be measured is positioned in the center of the first coil, while a magnetically tunable InSb photoconductive detector[36] is in the center of the second one. The InSb based cyclotron detector is a very sensitive detector with a spectral resolution on the order of 240 GHz (1 meV), over a spectral range from 0.8 THz up to 5 THz. The light emitted by the sample is guided toward the detector by a copper light pipe. The signal is detected as a voltage drop over the photoconductive detector, which is then amplified and measured via standard lock-in technique. The duty cycle (ratio of power-on to total time $T_0$) was set to a few percent (see Fig. SM9 c) to prevent overheating of the sample, which would damage the sample and distort the signal. In our experiments, the electric pulses had peak-to-peak values $V_A$ in the range of 10 to 100 V/cm and a duration $t_{on}$ of a few milliseconds. The spectrometer can be used in two different modes of use. The first one consists in keeping constant the magnetic field applied to the InSb detector, so as to fix the detection energy, while the magnetic field applied to the sample is swept. The second mode consists on the contrary in fixing the magnetic field on the sample side and sweeping that applied to the detector, so as to directly measure the emission spectrum. Experimental details can be found in [36].

## Data availability

Data are available upon reasonable request to the corresponding author. Dataset will also be uploaded on the Recherche Data Gouv repository (https://recherche.data.gouv.fr) once the manuscript is accepted and published.

## Methods only references

**Figure Legends/Captions**

**Fig. 1**: **Landau levels of Dirac fermions in graphene and HgTe QWs**. (a) spin-degenerate Landau levels in graphene with the arrows indicating the LL subsets equally spaced in energy. The inset schematically illustrates the linear band dispersion. (b,c) Landau levels in gapless and inverted HgTe QWs. The red and blue curves correspond to the eigenvalues $E_n^+$ and $E_n^-$, respectively. Here, the parameter C, representing the shift of the energy scale, is set to zero (see Eq. (5) in Supplemental Materials). The insets show the QW band structure. To stress the role of the quadratic terms, the dashed curves also represent the band structure at $\mathbb{B} = \mathbb{D} = 0$. The structure parameters used in the calculations are provided in Supplemental Materials.

**Fig. 2**: **Cyclotron emission observed on 8 nm HgTe QWs, measured on two different samples at T = 4 K.** a) and c) The signal is recorded with an *n*-InSb detector by fixing the magnetic field applied to the detector, i.e. the detection energy (right panels), while sweeping the magnetic field applied on the sample. b) and d) The signal is measured while fixing the magnetic field applied to the sample (values of which are seen on the right side of these panels), as a function of the detector's energy, i.e. swiping the magnetic field on the detector side.

**Fig. 3**: **Analysis of the cyclotron emission energies.** a) Energy of the cyclotron emission peaks as a function of the magnetic field for various samples. The symbols correspond to the maximum intensity of the CR peak. The dotted lines are linear fits according to the classical CR condition from which the corresponding cyclotron mass values were extracted. The light grey lines are the energies of intra-band Landau level transitions calculated for S8 sample. There are strikingly a very large number of possible optical transitions in the energy range at which the emission is measured. b) The Landau levels, from $N = 3$ to $N = 30$, calculated for S8 sample, as an example. The red curve represents the Fermi energy oscillations with magnetic field in the absence of disorder, which yields the zero broadening of Landau levels. To emphasize the semi-classical nature of the observed CR, the colored area shows the energy region in the vicinity of the Fermi energy, where transitions between the levels contribute to the cyclotron emission in sample S8 (schematically represented by arrows). The vertical dashed lines are several selected values of the filling factor to compare with the experimental results shown in the upper panel. c) The extracted cyclotron masses as a function of the carrier concentration of the samples in which the emission was measured. In this figure, the red and blue colors correspond to the QW thickness, 8 nm and critical, respectively. The solid curves represent the calculations based on Eq. (1). For comparison, we also provide the mass calculation within pure Dirac model involving only the linear terms i.e. at $\mathbb{B} = \mathbb{D} = 0$ (dashed lines). d) In order to highlight the tunability by concentration, we also represent experimental results by means of visualization of the energy and frequency emitted as a function of the density of carriers for two fixed magnetic fields, i.e. 0.5 Tesla and 1.0 Tesla. The shaded areas emphasize values extracted from the same sample with the two different operation modes, where full symbols represent again Mode 1 and open symbols Mode 2.

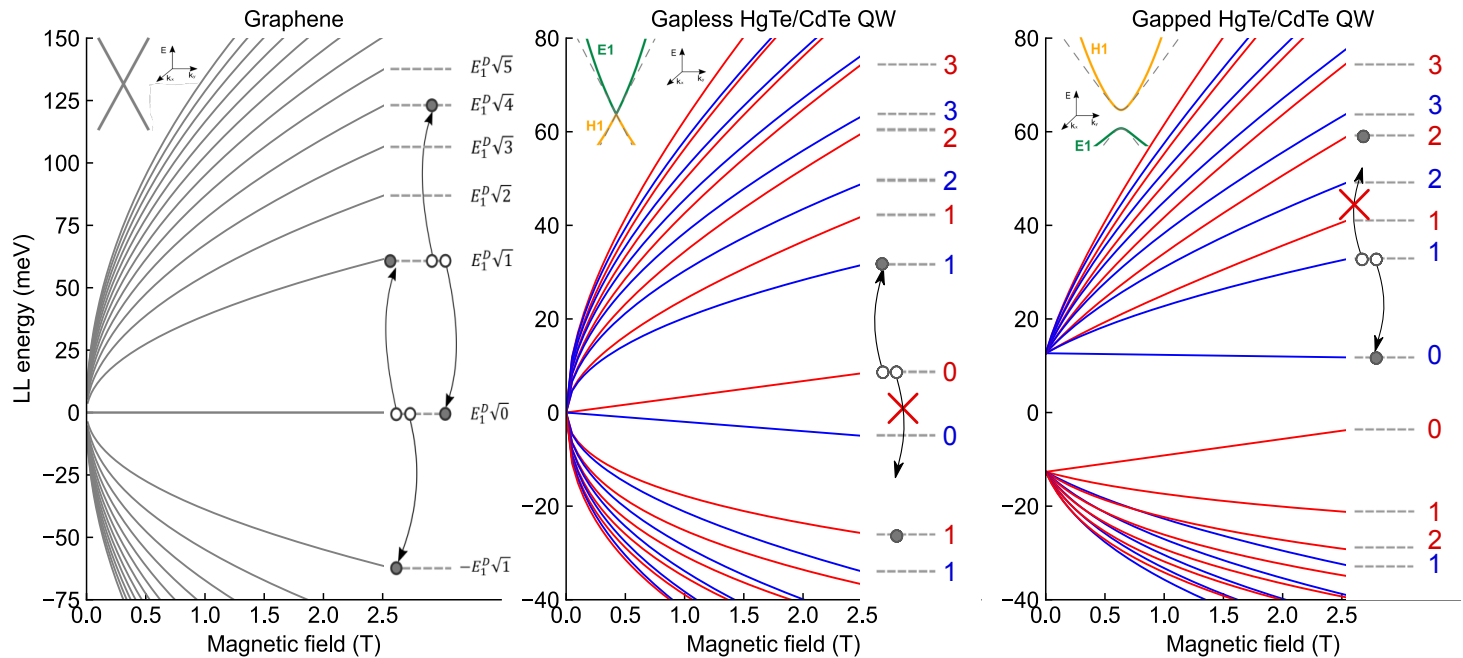

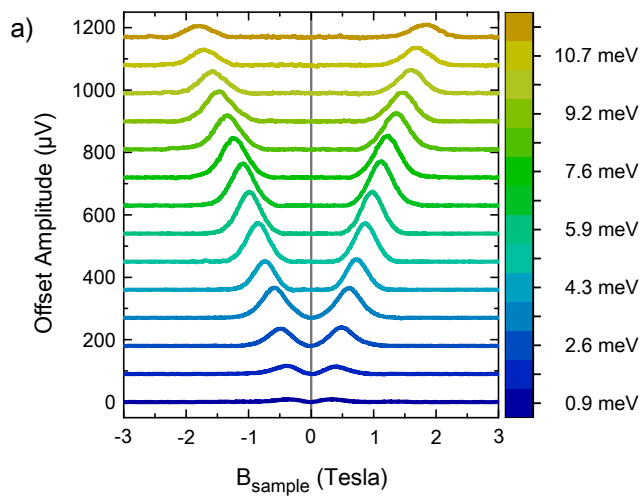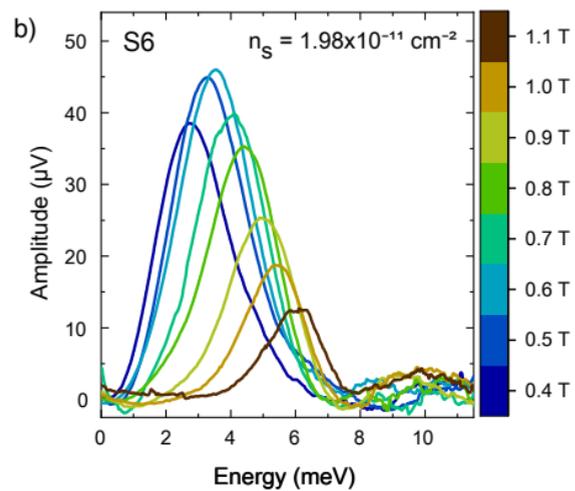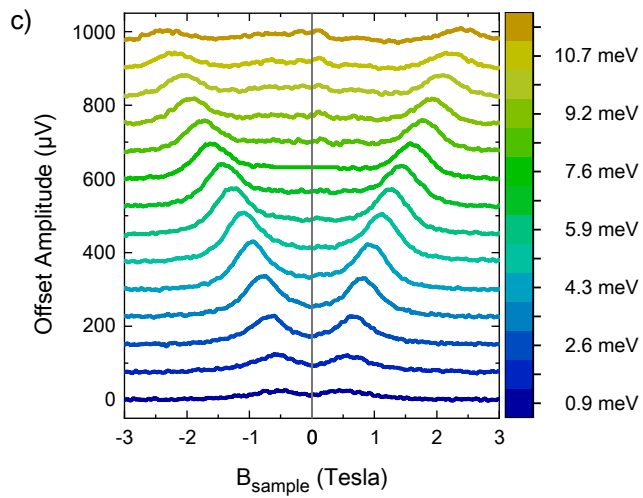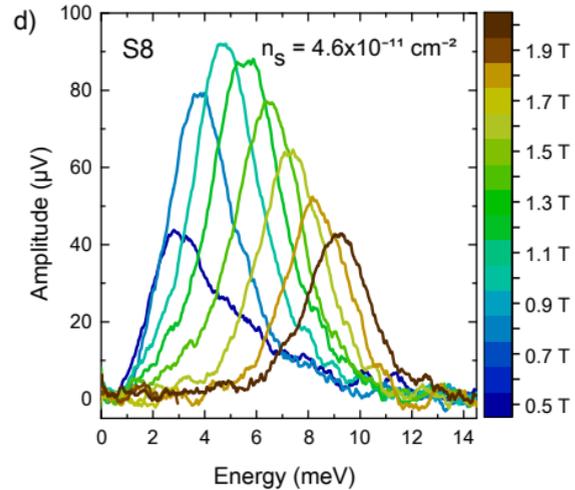

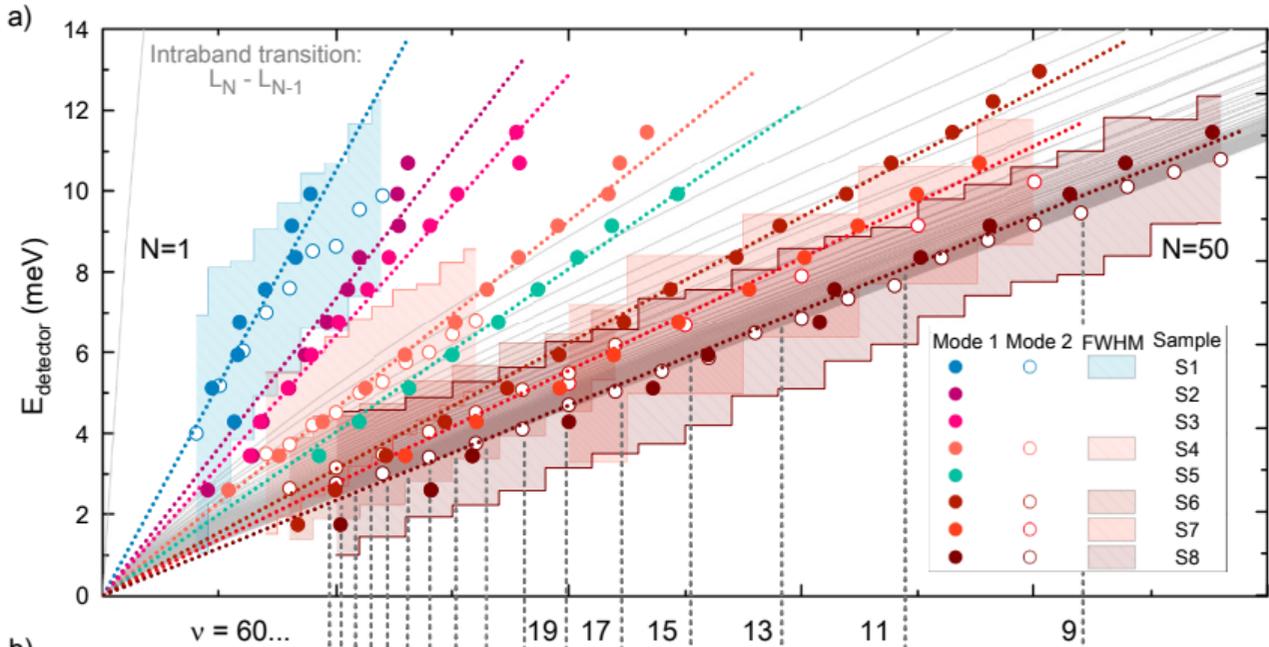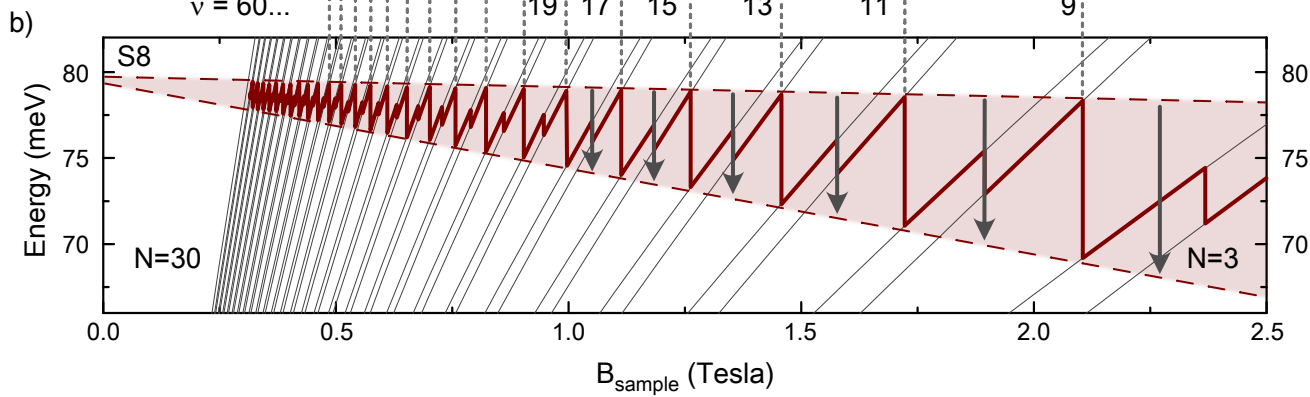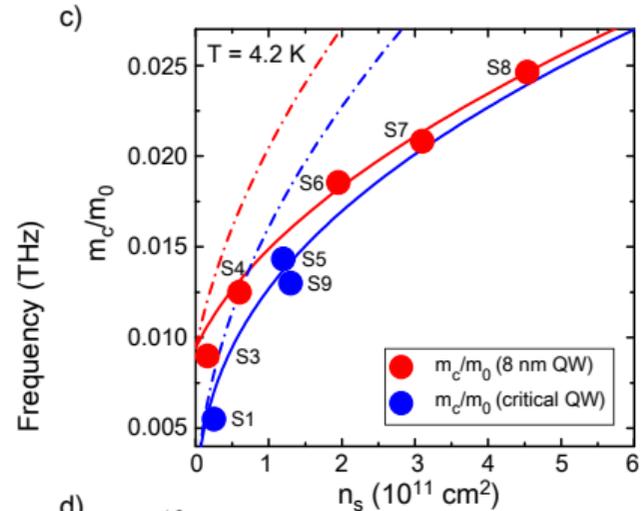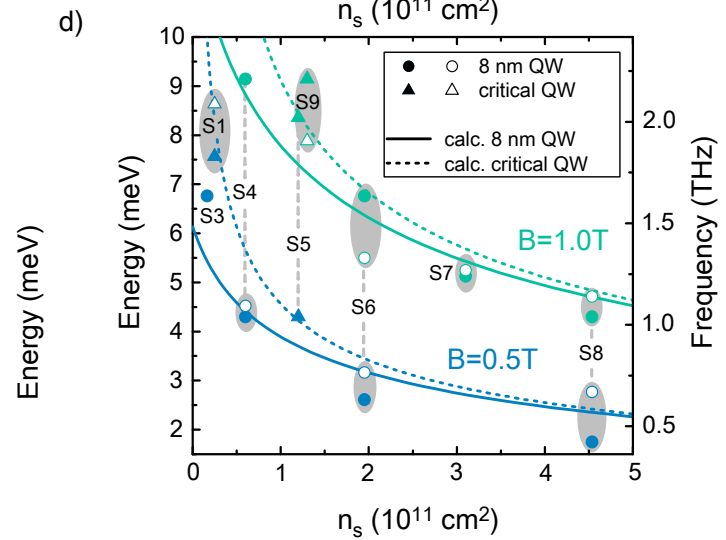